\documentclass{ifacconf}

\usepackage{graphicx}      %
\usepackage{natbib}        %

\usepackage{caption}
\usepackage{subcaption}
\usepackage{amssymb,amsmath,bm}
\usepackage{graphicx}
\newcommand{\beq}{\begin{equation}}
\newcommand{\eeq}{\end{equation}}
\begin{document}
\begin{frontmatter}
\vspace{-2em}
© 2024 Artur Wolek and James McMahon. This work has been accepted to IFAC for publication under a Creative Commons Licence CC-BY-NC-ND
\title{
Batch Estimation of a Steady, Uniform, Flow-Field from Ground Velocity and Heading Measurements} 

\thanks[footnoteinfo]{This work was  supported in part by NSF Grant No. 2301475.}

\author[First]{Artur Wolek} 
\author[Second]{James McMahon} 
\address[First]{Department of Mechanical Engineering and Engineering Science, University of North Carolina at Charlotte, Charlotte, NC, 28223 USA (e-mail: awolek@charlotte.edu).}
\address[Second]{Physical Acoustics Branch, Code 7135, Naval Research Laboratory, Washington, DC,   20375 USA  (e-mail: james.w.mcmahon8.civ@us.navy.mil)}

\begin{abstract}                %
This paper presents three batch estimation methods that use noisy ground velocity and heading measurements from a vehicle executing a circular orbit (or similar large heading change maneuver) to estimate the speed and direction of a steady, uniform, flow-field. The methods are based on a simple kinematic model of the vehicle's motion and use curve-fitting or nonlinear least-square optimization. A Monte Carlo simulation with randomized flow conditions is used to evaluate the batch estimation methods while varying the measurement noise of the data and the interval of unique heading traversed during the maneuver. The methods are also compared using experimental data obtained with a Bluefin-21 unmanned underwater vehicle performing a series of circular orbit maneuvers over a five hour period in a tide-driven flow.
\end{abstract}


\end{frontmatter}

\section{Introduction}
Autonomy and control algorithms can leverage estimates of the flow-field to improve the operation of marine vehicles, for example, in path planning to achieve greater efficiency and robustness. This paper is motivated by the authors' prior work \citep{wolek2021orbiting} that investigated time-optimal path planning for an underwater vehicle to  re-inspect points of interest along circular orbits with a sonar sensor. In such applications, the presence of a flow-field impacts the vehicle motion and the pointing angle of a directional sonar sensor (due to sideslip). The methods presented here can be used to estimate flow direction and magnitude on-the-fly (e.g., after executing each circular inspection orbit) to support path re-planning.

\begin{figure}[h!]
\centering
\includegraphics[width=0.45\textwidth]{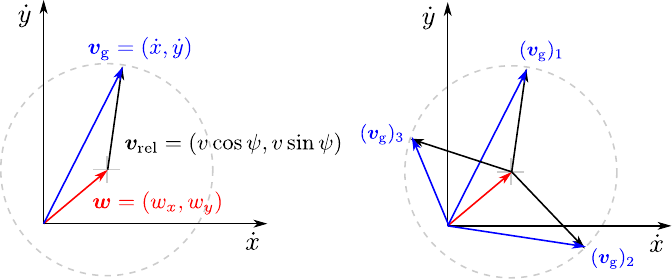}
\caption{Left: The velocity triangle. The dashed circle represents the set of attainable ground velocities for a fixed magnitude $||{\bm v}_{\rm rel}||$. The center of the circle is defined by the flow velocity ${\bm w}$. Right: At least three unique ground velocity vectors are required to determine the dashed-line circle.}
\label{fig:wind_triangle}
\end{figure}

For a vehicle operating in a steady, uniform, flow-field the  flow-relative velocity ${\bm v}_{\rm rel}$,  the flow velocity ${\bm w}$, and the inertial (ground) velocity ${\bm v}_{\rm g}$ are related by the equation ${\bm v}_{\rm g} = {\bm v}_{\rm rel} + {\bm w}$, as  illustrated in Fig.~\ref{fig:wind_triangle}. When two of the three quantities are known, the remaining one can be determined by direct computation. For marine vehicles, ground velocity, ${\bm v}_{\rm g}$, is  provided by the navigation system by processing GPS  position measurements, acoustic ranging measurements and/or a Doppler velocity log (DVL) sensor operating in bottom-lock mode. The vehicle flow-relative velocity ${\bm v}_{\rm rel}$ can also be measured with appropriate instrumentation (e.g., pitot tubes or a DVL operating in water current profiling mode).  
The flow velocity ${\bm w}$ may be inferred using various filtering techniques based on the vehicle's  kinematics \citep{rhudy2015aircraft}  or a dynamic model  \citep{hegrenaes2011model}.  Non-uniform flow-fields can be estimated, for example, using motion tomography \citep{chang2017motion}. 

The contributions of this paper are three batch estimation techniques to estimate  water current direction and magnitude for a marine craft using noisy measurements of kinematic variables and a basic model of the vehicle's motion. The methods do not require knowledge of the vehicle's flow-relative speed (since it is estimated as part of the procedure) and they are relatively simple to implement with minimal tuning required.  The first method is based on quadratic curve fitting to speed-over-ground magnitude and heading angle data when the vehicle executes a 360 degree heading change maneuver (e.g., a circular orbit). The second method relies on a constrained nonlinear  least-square optimization to match  speed-over-ground velocity components and heading data. The third method is similar to the second but uses  speed-over-ground magnitude, rather than velocity components, for optimization. The approaches are compared to each other, as well as to another existing method in the literature based on circular curve fitting, using both simulated and experimental data.

The remainder of the paper is organized as follows. Section~\ref{sec:curve_fitting} describes two curve-fitting based methods.  Section~\ref{sec:optimization_methods} describes two optimization-based methods.  Sections~\ref{sec:synthetic_data} and ~\ref{sec:experimental_data} provide a comparison with simulated and experimental data, respectively. Section~\ref{sec:conclusion} concludes the paper.

\section{Curve-Fitting-Based Methods}
\label{sec:curve_fitting}
Consider the following kinematic equations describing the motion of a marine craft in the horizontal plane:
\begin{align}
\dot x(t) &= f_x(\psi; v,  w, \theta) = v \cos \psi(t) + w \cos \theta \label{eq:xdot}\\
\dot y(t) &=  f_y(\psi; v , w, \theta ) =  v \sin \psi(t) + w \sin \theta\label{eq:ydot} \;,
\end{align}
where $(x,y)$ is the planar position, $v = ||{\bm v}_{\rm rel}||$ is the flow-relative speed,  $w = ||{\bm w}||$ is the magnitude of the current, and $\theta$ is the current direction. The  speed $v$ is assumed constant (e.g., corresponding to a fixed propeller speed). The current magnitude and direction, $(w,\theta)$, is assumed to be steady and uniform over the space in which a heading change maneuver is performed. 
This section describes two methods for estimating $(v, w, \theta)$  in \eqref{eq:xdot}--\eqref{eq:ydot}  based on curve fitting to noisy data of either  ground velocity components or ground velocity magnitude as a function of heading.
\subsection{Circular Curve Fit with $(\dot x, \dot y)$ Data}
In \citep{mclaren2008velocity} it was shown that given a minimum of three unique ground velocity component measurements, ${\bm v} = [\dot x, \dot y]^{\rm T}$ from  \eqref{eq:xdot}--\eqref{eq:ydot}, the parameters $(v, w, \theta)$ can be estimated. Here, we summarize this approach and use it for comparisons later on in Secs.~\ref{sec:synthetic_data} and \ref{sec:experimental_data}.

Consider the sketch in Fig.~\ref{fig:wind_triangle} (right panel) showing the  velocity triangle and a triplet of velocity triangles for three different flow-relative velocities.  A circle that passes through the three ground velocity vectors can be inscribed and its center corresponds to the current velocity ${\bm w} = [w_x, w_y]^{\rm T} = [w \cos \theta, w \sin\theta]^{\rm T}$. The radius of the circle is the vehicle's flow-relative speed $v$. 
One can confirm, by substituting \eqref{eq:xdot}--\eqref{eq:ydot}, that all points on the circle satisfy
\begin{align}
(\dot x - w_x) ^2 + (\dot y - w_y)^2  &= v^2  \;, 
\end{align}
which can re-arranged as
\begin{align}
 -  2 w_x \dot x    - 2  w_y \dot y  + (-v^2 + w_x^2  + w_y^2) &=  - \dot x^2  - \dot y^2 \label{eq:circle_constraint} \;.
\end{align}

\begin{figure}[h!]
\centering
\includegraphics[width = 0.3\textwidth]{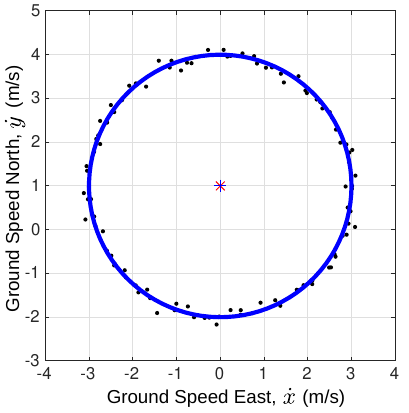}
\caption{Example of a circle fit (blue line) to a noisy $(\dot x, \dot y)$ dataset of $N = 100$ points (black dots) generated for parameters $(v, w, \theta) = $(3 m/s, 1 m/s, 90 deg.) with sensor noise standard deviation $\sigma = 0.1$. The red x marker indicates the true center point $(w_x, w_y)$. Using the circular curve-fit method the estimated values are $(\hat v, \hat w, \hat \theta)=$ (3.00 m/s, 0.991 m/s, 90.10 deg.). }
\label{eq:circle_example}
\end{figure}

Suppose that $N$ measurements $\{ \dot x_i , \dot y_i\}_{i=1}^N$ are recorded while the vehicle performs a maneuver.
In the absence of noise, the data satisfies ${\bm A}{\bm c} = {\bm b}$, where
\begin{align}
{\bm A} = 
\left[
\begin{array}{ccc}
\dot x_i & \dot y_i & 1 \\
\vdots & \vdots & \vdots \\
\dot x_N & \dot y_N & 1 \\
\end{array}
\right]\;, \qquad 
{\bm b}  = 
\left[
\begin{array}{c}
-\dot x_i^2 - \dot y_i^2 \\
\vdots \\
-\dot x_N^2 - \dot y_N^2 \\
\end{array}
\right]\;,  
\end{align}
and ${\bm c} =[ -2w_x, -2w_y, -v^2 + w_x^2 + w_y^2]^{\rm T} =  [c_1, c_2, c_3]^{\rm T}$. If, instead, the data $\{ \dot x_i , \dot y_i\}_{i=1}^N$ is corrupted by zero-mean additive Gaussian random variables with variance $\sigma^2$, then the system of equations can only be satisfied in a least-square sense (i.e., minimizing $J = || {\bm A}{\bm  c} - {\bm b} ||^2$) where ${\bm c}$ is estimated as ${\bm c} = {\bm A}^{\rm \dagger}{\bm b}$ and  ${\bm A}^\dagger = ({\bm A}^{\rm T}{\bm A})^{-1}$ is the matrix pseudo-inverse. The flow-field parameters are then:
\begin{align}
\hat w &= \frac{1}{2}\sqrt{c_1^2 + c_2^2} \\
\hat \theta &= {\rm atan2}(c_2, c_1) \\
\hat v &= \sqrt{ (1/4)(c_1^2 + c_2^2) - c_3}  \;.
\end{align}
An illustrative example of fitting a circular curve to $(\dot x, \dot y)$ data using this method is shown in Fig.~\ref{eq:circle_example} for a maneuver where the heading completes one full revolution (i.e., $\Delta \psi = 2\pi$). The approach can also be applied for maneuvers with $\Delta \psi  < 2 \pi$.

\subsection{Quadratic Curve Fit with $(v_{\rm g}, \psi)$ Data}
\begin{figure}
\centering
\includegraphics[width = 0.3\textwidth]{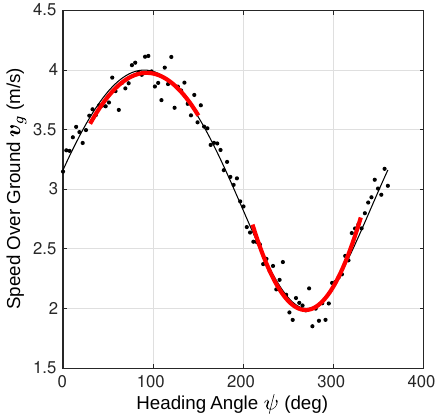}
\caption{Example of a quadratic curve fit  (red lines) to  $(v_{\rm g}, \psi)$ data (corresponding to the data in Fig.~\ref{eq:circle_example}).  The noise-free curve $v_{\rm g}(\psi)$ is shown as a black line. Using the quadratic curve-fit method the estimated values are $(\hat v, \hat w, \hat \theta) =$ (2.98 m/s, 0.994 m/s., 92.6 deg). }
\label{fig:qf}
\end{figure}

Now, suppose that the vehicle has access to the ground speed magnitude $v_{\rm g}  := ||{\bm v}_{\rm g}||$ (rather than the velocity components) along with heading angle information. That is, the data is $\{v_{{\rm g},i} , \psi_i \}_{i=1}^N$ where 
\begin{align}
v_{\rm g} = f_{v_{\rm g}}(\psi; v,  w, \theta)  &= \sqrt{\dot x(t)^2 + \dot y(t)^2} \\
&= \sqrt{ v^2  + w^2 + 2 v w \cos (\theta - \psi)  } \label{eq:v_g} \;.
\end{align}
Again, the data is assumed to be corrupted by additive zero-mean Gaussian noise. In general this assumption is made throughout this work, however, it is worth mentioning that for a vessel continuously rotating in one direction that is not aft-fore symmetric there may be a bias present based on the direction of rotation. To mitigate this bias, the dataset could include maneuvers performed in both clockwise and counter-clockwise orientations. 
If the heading undergoes a full revolution, then the curve $v_{\rm g}(\psi)$ contains a maximum and a minimum (see Fig.~\ref{fig:qf}). These minima and maxima can be approximated with a quadratic curve fit to estimate $(\hat w, \hat v, \hat \theta)$ as described next.

Differentiating \eqref{eq:v_g} with respect to $\psi$ gives the condition for a critical point:
\begin{align}
\frac{{\rm d}}{{\rm d}\psi } v_{\rm g} 
&= \frac{vw \sin (\theta - \psi) }{ \sqrt{ v^2  + w^2 + 2 v w \cos (\theta - \psi)  } }  = 0 \;,
\end{align}
which is satisfied at $\psi = \theta$ and $\psi = \theta + \pi$.  The speed over ground is a maximum when the vehicle heading is aligned with the current, $v_{\rm g}(\psi = \theta)  = v + w$, and a minimum when they are opposed, $v_{\rm g}(\psi = \theta + \pi)  = v -w $.
Now consider a Taylor series approximation at these two points. At the maximum:
\begin{align}
v_{\rm g} &\approx (v + w) - \frac{vw}{\sqrt{v^2 + 2 vw + w^2}}(\psi - \theta)^2 \\
&= \underbrace{ (-vw \delta^+) }_{a^+} \psi^2 + \underbrace{(2\theta vw \delta^+)}_{b^+} \psi   + \underbrace{(v + w - vw\delta^+ \theta^2)}_{c^+}  \;,
\end{align}
where $\delta^+ = 1/\sqrt{v^2 + 2 vw + w^2}$. Similarly, at the minimum
\begin{align}
v_{\rm g} &\approx (v - w) + \frac{vw}{\sqrt{v^2 - 2 vw + w^2}}(\psi - (\theta+ \pi))^2 \\
&=  \underbrace{(vw \delta^-)}_{a^-} \psi^2 \underbrace{- (2vw \delta^-(\theta +\pi))}_{b^-} \psi    \\ 
&\qquad \qquad + \underbrace{(v - w + vw\delta^- (\theta+\pi)^2)}_{c^-} \;,
\end{align}
where $\delta^- = 1/\sqrt{v^2- 2 vw + w^2}$. Since $\delta^+ < \delta^-$ the curvature at the minimum is greater.  Let $\psi_0^+$ denote a guess for the heading angle of the maximum.  Applying a second-order polynomial least-square fit to the the data on the interval $[\psi_0^+ - \lambda, \psi_0^+ + \lambda]$ to give the coefficients $(a^+, b^+, c^+)$. Equating these coefficients to the Taylor series approximation one can solve  for an estimate $(\hat v, \hat w, \hat \theta)$. A similar approach at the minimum gives the coefficients $(a^-, b^-, c^-)$ and another estimate. To utilize more of the available data,  both sets of parameters can be used. Let the coordinate of the maximum and minimum be
\begin{align}
(\psi_{\rm max}, v_{\rm max}) = ( -b^+/(2a^+),~ c^+ - (b^+)^2/(4a^+)) \\
(\psi_{\rm min}, v_{\rm min}) = ( -b^-/(2a^-),~ c^- - (b^-)^2/(4a^-)) \;,
\end{align}
respectively. 
Then, the estimates are:
\begin{align}
\hat v &= (v_{\rm max} + v_{\rm min})/2 \\
\hat w &= (v_{\rm max} - v_{\rm min})/2  \\
\hat \theta &= {\rm atan2}(\sin\psi_{\rm max} - \sin \psi_{\rm min},\cos \psi_{\rm max} - \cos \psi_{\rm min} )  \label{eq:hat_theta_qf}
\end{align}
The estimate $\hat \theta$ in \eqref{eq:hat_theta_qf} averages $\psi_{\rm min} - \pi$ and $\psi_{\rm max}$ while considering angle wrap-around. The approach applied to example data is shown in Fig.~\ref{fig:qf}.

\section{Optimization-Based Methods}
\label{sec:optimization_methods}
This section proposes two optimization-based approaches to estimate the vehicle speed and current magnitude and direction. The first method applies to $(\dot x, \dot y, \psi)$ data and the second method applies to $(v_{\rm g}, \psi)$ data.
\subsection{Least-square Optimization with $(\dot x, \dot y, \psi)$ Data}
Suppose that the vehicle obtains noise corrupted data of the form $\{\dot x_i, \dot y_i, \psi_i\}_{i=1}^N$ and consider the cost function:
\begin{align}
J &= \frac{1}{2}\sum_{i=1}^N\{  [\dot x_i - f_x(\psi_i; v,  w, \theta)  ]^2 +  [\dot y_i - f_y(\psi_i; v,  w, \theta)  ]^2 \} \notag \\
&=\frac{1}{2}\sum_{i=1}^N  \left\{  v^2 + w^2 + \dot x_i ^2 + \dot y_i ^2 - 2 \dot x_i  \cos \theta - 2 w \dot y_i \sin \theta   \right.\notag  \\
& \qquad  - 2 v \dot x_i\cos \psi_i  - 2 v \dot y_i \sin \psi_i + 2 v w \cos \theta \cos \psi_i \notag \\
& \qquad \left. + 2vw \sin \theta \sin \psi_i \right\} \;.
\label{eq:cost_xyh}
\end{align}
Now define the following constants that depend on the data collected:
\begin{align*}
k_1 := \sum_{i=1}^N  \dot x_i\;, \quad
k_2:= \sum_{i=1}^N  \dot y_i \;, \quad
k_3 :=   \sum_{i=1}^N  \dot x_i^2 \\
k_4:= \sum_{i=1}^N  \dot y_i^2\;, \quad
k_5:=   \sum_{i=1}^N \cos \psi_i \;, \quad
k_6 := \sum_{i=1}^N \sin \psi_i \\
k_7 :=  \sum_{i=1}^N \dot x_i \cos \psi_i \;, \quad
k_8 :=   \sum_{i=1}^N \dot y_i \sin \psi_i  \;.
\end{align*}
The cost function is then rewritten as
\begin{align}
J &= \frac{N}{2}^2v^2 + \frac{N}{2}w^2 + \frac{1}{2}k_3 +\frac{1}{2} k_4  - k_1 w \cos \theta \notag \\ 
& -k_2  w \sin \theta -  k_7 v  - k_8  v + k_5 vw \cos \theta  +  k_6 vw \sin \theta \;,
\label{eq:cost_k}
\end{align}
and notice that  \eqref{eq:cost_k} is linear in the data-dependent variables $k_1, \ldots ,k_8$. These terms only need to be computed once before the optimization proceeds. 
The gradient with respect to the parameters is:
\begin{align}
\nabla J = 
 \left[ 
\begin{array}{ccc}
{\partial J}/{\partial v},  & {\partial J}/{\partial w}, &  {\partial J}/{\partial \theta} 
\end{array}
\right]^{\rm T} \;,
\label{eq:jacobian}
\end{align}
where
\begin{align}
{\partial J}/{\partial v} &= Nv - k_7 - k_8 + k_5 w \cos \theta  + k_6 w \sin \theta \\
{\partial J}/{\partial w} 
&= Nw +(k_5 v-k_1) \cos \theta + (k_6 v - k_2 )\sin \theta   \\
{\partial J}/{\partial \theta} 
 &= (k_1  - k_5 v )  w\sin \theta + (k_6 v - k_2 )w \cos \theta \;.
\end{align}
The symmetric Hessian matrix is: 
\begin{equation}
{\bm H} 
=
\left[
\begin{array}{ccc}
H_{vv} & H_{vw} & H_{v\theta} \\
H_{wv} & H_{ww} & H_{w\theta} \\
H_{\theta v} & H_{\theta w} & H_{\theta\theta} 
\end{array}
\right]
\label{eq:Hessian}
\end{equation}
with entries
\begin{align*}
H_{vv} &= N \\
H_{vw} &=  k_5 \cos \theta + k_6 \sin \theta\\
H_{v\theta} &=  k_6 w \cos \theta -k_5 w \sin \theta\\
H_{ww} &= N\\
H_{w\theta}
&=  (k_1  - k_5 v) \sin \theta + (k_6 v - k_2) \cos \theta  \\
H_{\theta \theta} 
&= (k_1 - k_5 v)w \cos \theta+ (k_2 - k_6 v)w \sin \theta  
\end{align*}
 where $H_{wv} = H_{vw}$, $H_{\theta v} =  H_{v \theta}$, and $H_{\theta w}  = H_{ w \theta} $. 

\begin{figure}
\centering
\includegraphics[width = 0.35\textwidth]{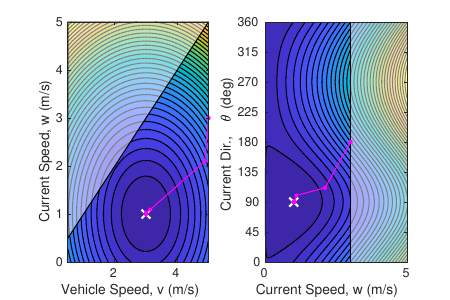}\\
\includegraphics[width = 0.35\textwidth]{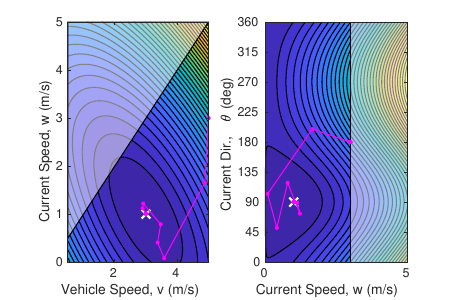}
\caption{Top: A surface plot of the cost function \eqref{eq:cost_xyh} over two slices of the $(v, w, \theta)$ parameter space for the noisy $(\dot x, \dot y, \psi)$ data corresponding to  Fig.~\ref{eq:circle_example}. The left panel is a slice of the cost function at the true value $\theta$, and the right panel is a slice at the true value $v$. The shaded areas represent the parameter space that does not satisfy the constraints. The true $(v,w, \theta)$ point is indicated by a white x marker, and the optimizer iterations, projected on to each plane, are shown in magenta. Bottom:  Visualization of the same optimization using data generated for a maneuver with a smaller heading angle change of $\Delta \psi = \pi$. }
\label{fig:opt_XYP}
\end{figure}

The cost, gradient, and Hessian are essential for optimization (e.g., via Newton-Rhapson's method). By deriving the analytical forms of the gradient and Hessian  the use of finite difference techniques can be avoided to improve optimization efficiency and robustness. In this work, we supply $\nabla J$ and $H$ to an interior-point optimizer implemented by the \texttt{fmincon} function in MATLAB.  A constrained optimization problem is formulated subject to the linear inequality constraints  
\begin{equation}
\left[
\begin{array}{ccc}
-1 & 0 & 0 \\
1 & 0 & 0 \\
0  & 1 & 0 \\
-1 & 1 & 0 \\
0 & 0 & 1 \\
0 & 0 & -1 \\
\end{array}
\right]
\left[ 
\begin{array}{c}
v \\
w \\
\theta 
\end{array}
\right ]
\leq 
\left[ 
\begin{array}{c}
v_{\rm min} \\
v_{\rm max} \\
0 \\
0 \\
0 \\
2\pi  \\
\end{array}
\right ] \;,
\label{eq:constraints}
\end{equation}
where $v_{\rm min}$ and $v_{\rm max}$ are user-supplied lower and upper bounds on the vehicle's flow-relative speed. The constraints \eqref{eq:constraints} also encode for the assumption $w \leq v$ that is required for the problem to be well posed.  Local minima can occur along the boundaries of the constraint set since the topology of $\theta$ is not considered in \eqref{eq:constraints}. Additionally, local minima may occur  with noisier datasets or when the maneuver contains a smaller range of heading angles.  To improve robustness the optimization is  run from multiple start points and the lowest cost solution is selected. The initial starting points are selected as the corners of the polytope represented by the constraints and as the centroid of the polytope. %
An example of the cost function and optimization scheme is shown in Fig.~\ref{fig:opt_XYP}.  The level sets of the cost function are indicative of the uncertainty in the parameter space around the estimated point.
\subsection{Least-square Optimization with $(v_{\rm g}, \psi)$ Data}
Lastly, we consider an optimization-based estimation algorithm that assumes speed-over-ground measurements and heading data are available. Consider the cost function
\begin{align}
J &= \frac{1}{2}\sum_{i=1}^N [(v_{\rm g})_i - f_{v_{\rm g}}(\psi; v,  w, \theta)  ]^2  \;.
\label{eq:cost_vg}
\end{align}
The gradient with respect to the parameters has the same form as \eqref{eq:jacobian} with entries
\begin{align}
\frac{\partial J}{\partial v} 
&= -\sum_{i=1}^N
\frac{\left(2 v+2 w (c_{\psi\theta})_i\right) \left((v_{\rm g})_i-\sqrt{ \alpha_i}\right)}{\sqrt{ \alpha_i}} \\
\frac{\partial J}{\partial w}  
&= -  \sum_{i=1}^N\frac{\left(2 w+2 v (c_{\psi\theta})_i\right) \left((v_{\rm g})_i-\sqrt{ \alpha_i}\right)}{\sqrt{ \alpha_i}}\\
\frac{\partial J}{\partial \theta} 
&= -  \sum_{i=1}^N \frac{2 v w (s_{\psi\theta})_i \left((v_{\rm g})_i-\sqrt{ \alpha_i}\right)}{\sqrt{ \alpha_i}} \;, 
\label{eq:grad2}
\end{align}
where $\alpha_i =  \left(v^2+2\cos\left(\psi_i-   \theta\right)v w+w^2\right)$. 
The symmetric Hessian matrix is the same as in \eqref{eq:Hessian} 
with entries
\begin{align*}
H_{vv} &= \sum_{i=1}^N \frac{2 \left({\alpha_i}^{3/2}-(v_{\rm g})_i w^2+(v_{\rm g})_i w^2 {(c_{\psi \theta})_i}^2\right)}{{\alpha_i}^{3/2}} 
\end{align*}
\begin{align*}
H_{vw} &= \sum_{i=1}^N \frac{\left(2 v+2 w (c_{\psi \theta})_i\right) \left(2 w+2 v (c_{\psi \theta})_i\right)}{2 \alpha_i }  \\
& \quad 
-\frac{2 (c_{\psi \theta})_i \left((v_{\rm g})_i-\sqrt{\alpha_i}\right)}{\sqrt{\alpha_i}} \\ 
& \quad +\frac{\left(2 v+2 w (c_{\psi \theta})_i\right) \left(2 w+2 v (c_{\psi \theta})_i\right) \left((v_{\rm g})_i-\sqrt{\alpha_i}\right)}{2 {\alpha_i}^{3/2}} 
\end{align*}
\begin{align*}
H_{v\theta} &= \sum_{i=1}^N -\frac{2 w (s_{\psi \theta})_i \left((v_{\rm g})_i w^2-{\alpha_i}^{3/2}+v (v_{\rm g})_i w (c_{\psi \theta})_i\right)}{{\alpha_i}^{3/2}}
\end{align*}
\begin{align*}
H_{ww} &=\sum_{i=1}^N  -\frac{2 v (s_{\psi \theta})_i \left(v^2 (v_{\rm g})_i-{\alpha_i}^{3/2}+v (v_{\rm g})_i w (c_{\psi \theta})_i\right)}{{\alpha_i}^{3/2}} 
\end{align*}
\begin{align*}
H_{w\theta}&=\sum_{i=1}^N  -\frac{2 v (s_{\psi \theta})_i \left(v^2 (v_{\rm g})_i-{\alpha_i}^{3/2}+v (v_{\rm g})_i w (c_{\psi \theta})_i\right)}{{\alpha_i}^{3/2}}
\end{align*}
\begin{align*}
H_{\theta \theta} 
&=\sum_{i=1}^N  -\frac{2 w (s_{\psi \theta})_i \left((v_{\rm g})_i w^2-{\alpha_i}^{3/2}+v (v_{\rm g})_i w (c_{\psi \theta})_i\right)}{{\alpha_i}^{3/2}}
\end{align*}
where $c_{\psi\theta} = \cos\left(\psi_i-\theta\right)$ and $s_{\psi\theta} = \sin\left(\psi_i-\theta\right)$ and with $H_{wv} = H_{vw}$, $H_{\theta v} =  H_{v \theta}$, and $H_{\theta w}  = H_{ w \theta} $.  Both the gradient and the Hessian are well-defined only if $\alpha_i \neq 0$. In practice, the case $\alpha_i = 0$ occurs rarely and during implementation we check for and remove any such instances from the dataset. 

\begin{figure}
\centering
\includegraphics[width = 0.35\textwidth]{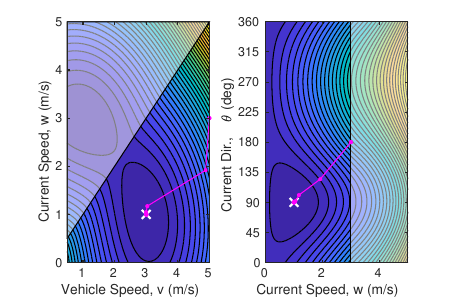}\\
\includegraphics[width = 0.35\textwidth]{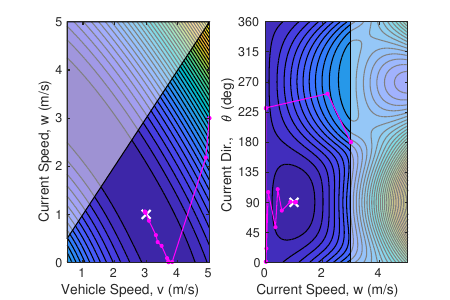}
\caption{Top: A surface plot of the cost function \eqref{eq:cost_vg} for the dataset in Fig.~\ref{fig:qf} with optimization iterations shown in magenta. Bottom: A similar plot for a dataset for a smaller change in heading $\Delta \psi = \pi$. Refer to the Fig.~\ref{fig:opt_XYP} caption for further details.}
\label{fig:opt_VP}
\end{figure}

Unlike the previous optimization case, the cost function, gradient, and Hessian depend non-linearly on the data. The summations involving $\{\dot x_i, \dot y_i, \psi_i\}_{i=1}^N$ must be recomputed at each iteration.  As described earlier, the analytical forms of $\Delta J$ and $H$ are provided to \texttt{fmincon} along with constraints \eqref{eq:constraints}. An example of the cost function and optimization sequence is shown in Fig.~\ref{fig:opt_VP}.

\section{Comparison using Simulated Data}
\label{sec:synthetic_data}
\begin{figure}[h!]
\centering
\includegraphics[width=0.43\textwidth]{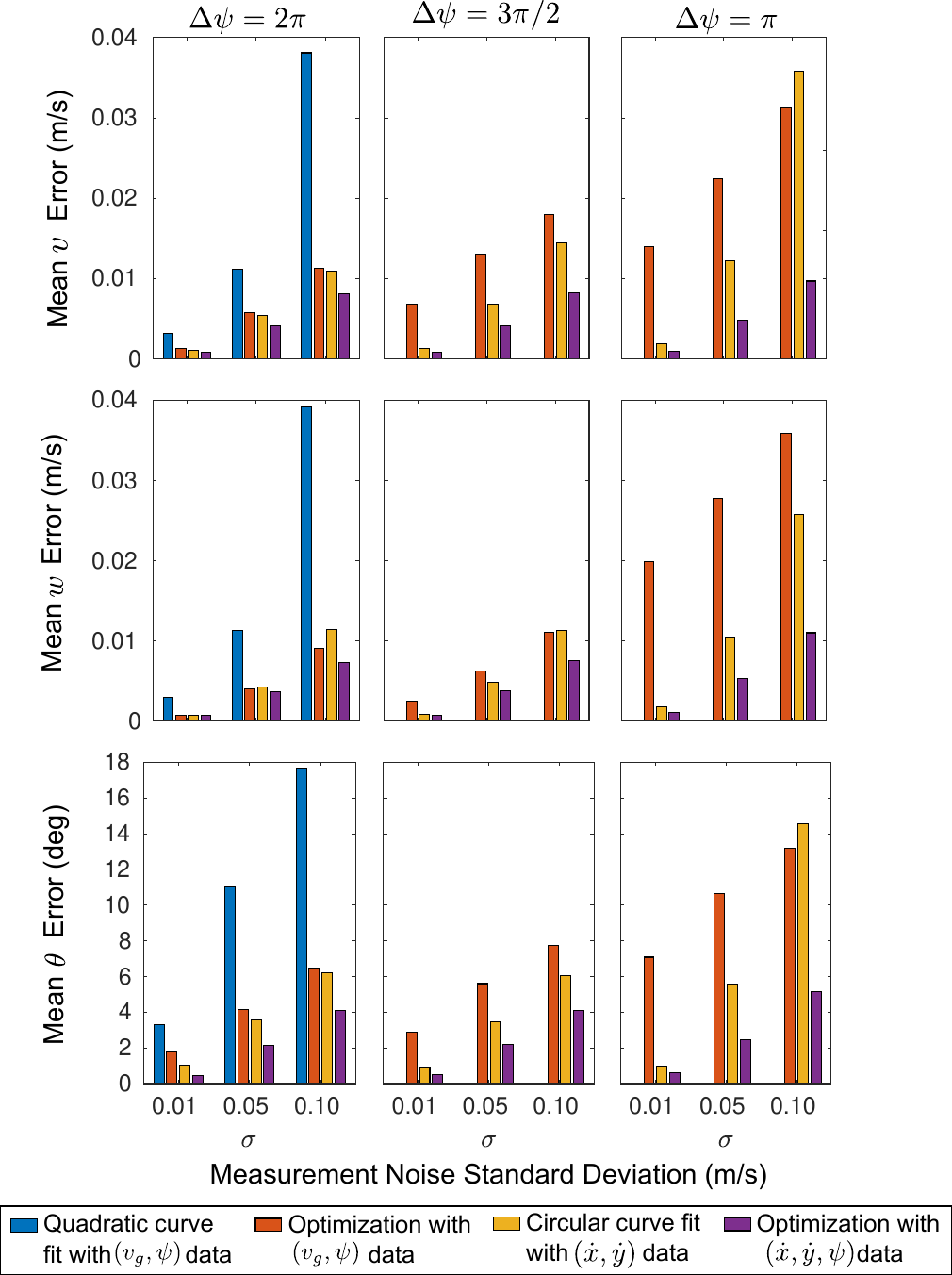}
\caption{Monte Carlo experiment results. Mean error in $v$, $w$, and $\theta$ is presented on each row with columns corresponding to a different heading change intervals. } 
\label{fig:monte_carlo}
\end{figure}

To compare the performance of the four methods described in Secs. \ref{sec:curve_fitting} and \ref{sec:optimization_methods}, a Monte Carlo experiment was conducted using simulated data. The data was simulated for three different noise settings, $\sigma = \{ 0.01, 0.05, 0.10 \}$ m/s, and three different heading angle changes, $\Delta \psi = \{ 2 \pi, 3 \pi / 2, \pi \}$. For each combination of $\sigma $ and $\Delta \psi$ a total of 250 unique datasets of $N = 100$ noisy samples were generated and transformed to the appropriate form for each algorithm. Each dataset was generated by randomizing the true parameters $(v, w, \theta)$ and drawing them uniformly at random from the interior of the polytope represented by the constraints with $v_{\rm min} = 0.5$ and $v_{\rm max} = 5$ m/s. For each dataset the circular fit algorithm, optimization with $(\dot x, \dot y)$ data, and optimization with $(\dot v_g, \psi)$ data were tested.  The quadratic fit approach was evaluated only for the  $\Delta \psi = 2\pi$ case.  The estimate each algorithm produced was compared to the actual value and the absolute error was recorded. The mean of the absolute error, average over 250 trials, is shown in Fig.~\ref{fig:monte_carlo}. 

As expected, the error for each algorithm increased as the measurement noise is increased and as the range of data is reduced. The quadratic fit method produced the largest errors (e.g., almost 18 deg in mean current direction error for the highest noise setting). The optimization with the $(v_{\rm g}, \psi)$ data was not as accurate as the optimization with $(\dot x, \dot y, \psi)$ data. This is not surprising since more information is  available when the components of the ground velocity are known, rather than the magnitude. The circle fit, which uses $(\dot x, \dot y)$ data, in most cases had an intermediate error in comparison to the the two optimization-based methods. The optimization method using $(\dot x, \dot y, \psi)$ data produced the most accurate results --- for example, in the case of data collected for $\Delta \psi  = 2\pi$ maneuvers the vehicle and current speed errors were $<$ 0.01 m/s and the current direction error was $<$ 5 degrees. 

\section{Comparison using Experimental Data}
\label{sec:experimental_data}
\begin{figure}
\centering
\includegraphics[width=0.32\textwidth]{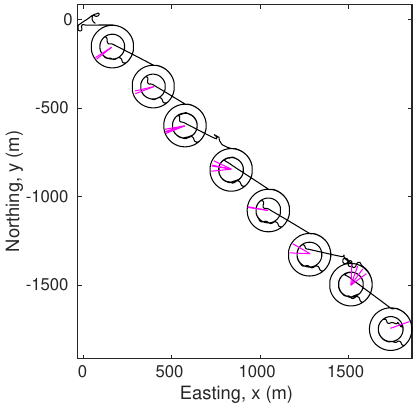}
\caption{Ground tracks during an at-sea experiment with overlay of estimated flow-field direction after each circular orbit (magenta lines).}
\label{fig:example_tracks}
\end{figure}

\begin{figure}
\centering
\includegraphics[width=0.4\textwidth]{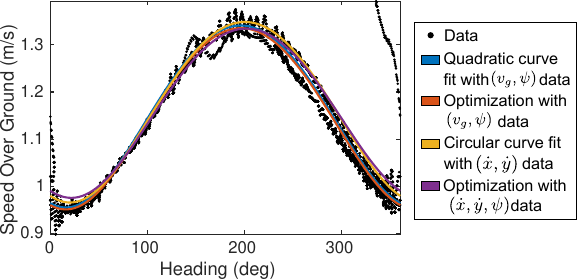}
\caption{Example experimental data along a single orbit with overlayed $v_{\rm g}(\psi)$ curves for each estimate. }
\label{fig:example_boston}
\end{figure}

The four methods described in Secs.~\ref{sec:curve_fitting}  and \ref{sec:optimization_methods} were also evaluated using experimental data obtained by a Bluefin-21  unmanned underwater vehicle operated by the U.S. Naval Research Laboratory. The Bluefin-21 is a 20 ft length, 21 inch diameter vehicle developed by Bluefin Robotics that features a fiber-optic gyroscope inertial navigation system and Doppler velocity log (DVL) navigation suite. The vehicle was deployed on June 22, 2016 near Boston Harbor southeast of Nahant Bay. The vehicle was programmed to execute a series of circular orbits around points of interest to test an onboard sonar system. The ground track of the vehicle is shown in Fig.~\ref{fig:example_tracks} along with the flow direction estimated  by the optimization method with $(\dot x, \dot y, \psi)$ data. 
The radii of the smaller and larger orbits were 70 and 120 meters, respectively. The data under analysis was truncated to be within 10 meters of these orbits and input into all four estimation algorithms. An example of the dataset along one such orbit is shown in Fig.~\ref{fig:example_boston}. The parameters estimated by each algorithm were used to superimpose a $v_{\rm g}(\psi)$ curve over the data points. The dataset exhibits outliers that do not conform to the model (e.g., due to transient motions as the vehicle enters and exits each orbit). To address this issue the optimization algorithms are wrapped in a random sample consensus (RANSAC) algorithm outer loop.

The estimated current direction and magnitude was also compared to historical data from an instrumented buoy (identifier: BOS1132) deployed by the National Oceanic and Atmospheric Administration (NOAA) located in Stellwagen Bank about 51 km away directly east from the testing area (15nm NNE of Race Point at a depth  of 27ft). The results show good agreement with the ebb and flow tidal directions reported by the buoy. The current magnitude matches the general trend and variations may be due to the spatial separation of the buoy and vehicle leading to different tidal flows closer to shore.

\begin{figure}
\centering
\includegraphics[width=0.46\textwidth]{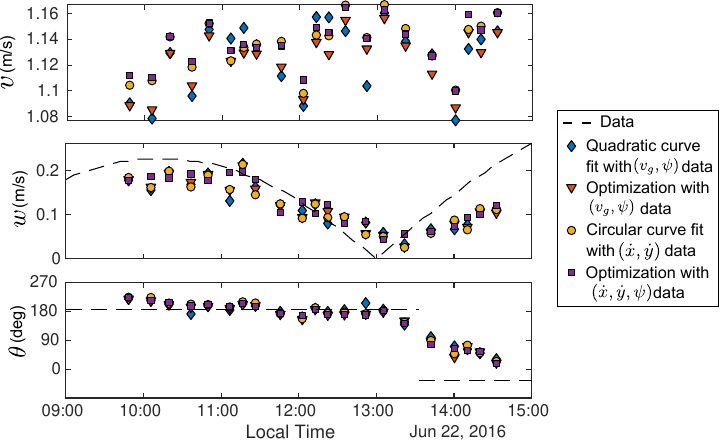}
\caption{Comparison of estimaties with NOAA buoy data. }
\label{fig:noaa}
\end{figure}

\section{Conclusion}
\label{sec:conclusion}
Three batch estimation methods were presented that determine the direction and magnitude of a steady, uniform, flow-field, and the vehicle's speed, using noisy kinematic measurements during heading change  maneuvers, such as circular orbits or 180 degree turns. The three methods proposed included a quadratic curve fitting approach with $(v_{\rm g}, \psi)$ data and least-square optimization methods with either $(\dot x_, \dot y, \psi)$ or $(v_{\rm g}, \psi)$ data. The methods were compared through a Monte Carlo experiment with simulated data to illustrate the impact of measurement noise and heading angle change during the maneuver. The comparison included an existing circular curve fitting method from the literature that uses $(\dot x, \dot y)$ data.  The results indicated that the optimization with $(\dot x_, \dot y, \psi)$ led to the lowest mean errors. The methods were also evaluated using a experimental data obtained by an underwater vehicle performing a series of circular orbits during a tidal change.  The estimated values had modest agreement with data recorded by a nearby buoy. 

The advantages of the proposed methods are that they are relatively simple to implement, depend on only a small number of parameters, and do not require a vehicle model or assumption of vehicle flow-relative speed. The optimization methods also allow visualizing the uncertainty in the parameter space as level sets of a corresponding cost function. Future work may consider running the estimators onboard a vehicle, comparing to dynamic state estimators (e.g., Kalman filters that estimate flow-field conditions), and using flow-field estimates to optimize mission plans.
\bibliographystyle{unsrt}
\bibliography{refs} 
\end{document}